\documentclass[11pt, letterpaper]{article}

\usepackage[letterpaper, margin=1in]{geometry} 
\usepackage{layout} 

\usepackage[bookmarksnumbered,bookmarksopen,colorlinks,citecolor=blue,linkcolor=blue]{hyperref}

\usepackage{natbib}

\usepackage{amsfonts}
\usepackage{mathrsfs}
\usepackage{amsmath}
\usepackage{amssymb}
\usepackage{amsthm}

\usepackage{color}
\usepackage{graphicx}

\usepackage{ifthen}
\usepackage{authblk}

\usepackage{pdflscape}

\newcommand{\Keywords}[1]{\par\noindent{\small{\em Keywords\/}: #1}}

\begin{document}

\title{Novel decision-theoretic and risk-stratification metrics of predictive performance: Application to deciding who should undergo genetic testing}

\author[1]{Hormuzd A. Katki}
\affil[1]{Division of Cancer Epidemiology and Genetics, National Cancer Institute,Bethesda, MD,	USA.}
\maketitle

\setcounter{page}{1}   
\pagenumbering{roman}   

\setcounter{tocdepth}{3} 


\begin{abstract}
	Currently, women are referred for \textit{BRCA1/2} mutation-testing only if their family-history of breast/ovarian cancer implies that their risk of carrying a mutation exceeds 10\%.  However, as mutation-testing costs fall, prominent voices have called for testing all women, which would strain clinical resources by testing millions of women, almost all of whom will test negative.  To better evaluate risk-thresholds for \textit{BRCA1/2} testing, we introduce two broadly applicable, linked metrics: Mean Risk Stratification (MRS) and a decision-theoretic metric, Net Benefit of Information (NBI).  MRS and NBI provide a range of risk thresholds at which a marker/model is "optimally informative", in the sense of maximizing both MRS and NBI.  NBI is a function of only MRS and the risk-threshold for action, connecting decision-theory to risk-stratification and providing a decision-theoretic rationale for MRS.  AUC and Youden's index reflect on both the fraction of maximum MRS, and of maximum NBI, attained by the marker/model, providing AUC and Youden's index with long-sought decision-theoretic and risk-stratification rationale.  To evaluate risk-thresholds for \textit{BRCA1/2} testing, we propose an eclectic approach considering AUC, Net Benefit, and MRS/NBI. MRS/NBI interpret AUC in the context of mutation-prevalence and provide a range of risk thresholds for which the risk model is optimally informative.
\end{abstract}

\Keywords{
	AUC; \textit{BRCA1} and \textit{BRCA2}; Diagnostic Testing, Mean Risk Stratification, Net Benefit; Net Benefit of Information; risk prediction ; ROC; Screening; Youden's index.
}

\setcounter{page}{1}    
\pagenumbering{arabic}  


\section{Introduction}
\label{s:intro}

There is a torrent of new markers and risk-prediction models, but unfortunately, historically used metrics for markers and risk models provide at best indirect information about utility, or even predictiveness, for clinical/public-health use.  The odds-ratio is a well-known poor measure of predictiveness~\citep{Kraemer2004,Pepe2004}.  When comparing two tests, it is uncommon for one test to have both higher sensitivity and specificity, or both higher positive predictive value (PPV) and lower complement of the negative predictive value (cNPV). Two commonly used summary statistics, Youden's Index~\citep{YOUDEN1950} and the Area Under the Receiver Operating Characteristic Curve (AUC)~\citep{Hanley1982}, have been correctly criticized for not taking predictive values (i.e. absolute risks) into account, and for not permitting differential weighting of false-positive versus false-negative errors~\citep{GREENHOUSE1950,Hilden1991}.  In spite of the well-known criticisms, the AUC remains by far the most popular metric in scientific practice.

There has been much research on improved metrics, such as risk-reclassification metrics~\citep{Cook2007,Pencina2008}, and especially metrics with decision-theoretic justification~\citep{Gail2005,Baker2009a}.  Decision-theoretic metrics often rely only on specifying a risk-threshold~\citep{Pauker1980} that implicitly accounts for benefits, harms, and costs.  The most popular metric is the Net Benefit~\citep{Vickers2006}, which provides a range of risk-thresholds, below which action should always be taken, above which action should never be taken, and within which using the marker/model provides the most utility.  Net Benefit is an important advance, but we feel that its meaning as an abstract "utility" measure is hard for scientists to interpret.  

A parallel line of research focuses on "risk stratification", a ubiquitous term in medicine (20,684 papers in PubMed on August 18 2017).  Although "risk stratification" is a broadly used term, we define it as the ability of a marker/model to separate those at high risk of disease from those at low risk~\citep{Wentzensen2013}.  "Risk stratification", as a term, is much less common in the statistical literature (75 papers in Scopus on August 18 2017).  Important advances in risk-stratification include the predictiveness curve~\citep{Copas1999,Huang2007} and its summary metric Total Gain~\citep{bura2001binary}.

We introduce two new broadly applicable, linked metrics that connect the decision-theoretic and  risk-stratification approaches to evaluating markers/models.  We first define Mean Risk Stratification (MRS) as the average change in risk of disease (posttest-pretest) revealed by the marker/model.  MRS is twice the cross-product \textit{difference} of the joint probabilities in a 2x2 table, which is easily remembered by analogy with odds ratios.  We then define a decision-theoretic metric, the Net Benefit of Information (NBI): the increase in expected utility from using the marker/model to select people for intervention \textit{versus} randomly selecting people for intervention.  Our 4 key results are:
\begin{enumerate}
	\item Youden's index and AUC reflect on (1) the fraction of the maximum possible risk-stratification (i.e. MRS) that is attained by the marker/model, and (2) the fraction of the maximum possible utility gain over random selection (i.e. NBI) that is attained by the marker/model.  MRS and NBI provide Youden's index and AUC with risk-stratification and decision-theoretic justification, the lacks of which have long been a criticism of Youden's index and AUC.
	
	\item For rare diseases, high AUC does not imply high risk-stratification or utility gain over random selection.  AUC must be considered in light of disease prevalence, which is automatically done by MRS and NBI.
	
	\item NBI is a function of only MRS and the risk-threshold for action, connecting decision-theory to risk-stratification and providing a decision-theoretic rationale for MRS.
	
	\item MRS and NBI provide a range of risk thresholds for which the marker/model is "optimally informative", in the sense that the risk-thresholds maximize both risk-stratification and the utility gain over random selection.
\end{enumerate}

We apply MRS and NBI to the controversy over who should get tested for mutations in the \textit{BRCA1/2} genes, which cause high risks of breast and ovarian cancer~\citep{Kuchenbaecker2017}.  The mutations are rare in the general population ($\approx0.25\%$), but are 10 times more common among Ashkenazi-Jews~\citep{STRUEWING1997}.  Currently, according to the UK National Institute for Health and Care Excellence (NICE) and the US Preventive Services Task Force, women are referred for mutation testing only if they have a strong family history of breast or ovarian cancer~\citep{Moyer2014} as quantified by a risk model calculating that their risk of carrying a mutation exceeds 10\%~\citep{NICE2017}.  However, as mutation-testing costs fall, prominent voices have called for testing \textit{all} women~\citep{King2014,GenomeWeb2017}.  \textit{BRCA1/2} testing is already being offered to a large unselected Canadian population as a demonstration project~\citep{GenomeWeb2017a}.  Testing all women would strain clinical resources by testing millions of women, $99.75\%$ of whom will test negative.  At US \$500-\$1,000 per test, testing millions of women has clear commercial implications.  

In contrast, we recently demonstrated that 80\% of Ashkenazi-Jewish mutation-carriers could be identified by testing only 44\% of Ashkenazi-Jewish women~\citep{Best2017}.  This is achieved by a low mutation-risk threshold of 0.78\%, far lower than the current 10\% UK NICE and US recommendation.   However, we could not formally justify any choice of risk-threshold.  We seek to better understand the implications of different choices of risk threshold.  Our eclectic approach considers multiple metrics, including AUC, Net Benefit, and MRS/NBI.  The ranges of useful risk-thresholds, as determined by Youden's index, AUC, MRS, NBI, and Net Benefit, always overlap at risk threshold equaling disease prevalence. The value of MRS/NBI is to interpret AUC in the context of prevalence and to provide a range of risk thresholds for which the risk model is optimally informative.  \href{http://analysistools.nci.nih.gov/biomarkerTools}{Our MRS webtool is available.}

\section{Mean Risk Stratification (MRS)}
\label{sec:mean-risk-strat}

Because continuous markers or risk models require dichotomization to determine action, we refer to a marker or model $M$, dichotomized at a cutpoint $m_0$, as a \emph{test}: $M+=\{M \ge m_0\}$ is positive and $M-=\{M<m_0\}$ is negative.   In the absence of test results or other pretest information, each individual can only be assigned as a best guess the same population-average risk $P(D+)$.  Upon taking the test, 2 outcomes are possible:
\begin{enumerate}
	\item With probability $P(M+)$, the test is positive.  The person's risk increases from $P(D+)$ to Positive Predictive Value ($PPV=P(D+|M+)$), an increase of $PPV-P(D+)$.
	\item With probability $P(M-)$, the test is negative.  The person's risk decreases from $P(D+)$ to complement of Negative Predictive Value: $cNPV=P(D+|M-)$.  The person's risk decreases by $P(D+)-cNPV$.
\end{enumerate}    
Mean Risk Stratification (MRS) is a weighted average of the increase in risk among those who test positive  and the decrease in risk among those who test negative:
\begin{eqnarray}
\label{eq:1}
MRS(m_0) = \{PPV-P(D+)\}P(M+) + \{P(D+)-cNPV\}P(M-)
\end{eqnarray}
MRS is the average difference between predicted post-test individual risk (either PPV or cNPV) and pretest population-average risk $P(D+)$.  Stated simply, MRS is the average change in risk that a test reveals.  For instance, an MRS of 6\% means that if a person uses this test, the person learns that their disease risk will increase or decrease by an average of 6 disease cases per 100 people.  

Importantly, via the definition of $M+$ and $M-$, MRS is a function of the cutpoint $m_0$ that defines  risk thresholds.  We will plot MRS for all possible $m_0$ to gain insight on the value of possible risk thresholds.  The cutpoint $m_0$ that maximizes MRS is always where the risk threshold equals disease prevalence (see Appendix).  Thus, the range of risk thresholds where MRS is highest will always contain disease prevalence.  

Total Gain is the special case of MRS where cutpoint $m_0$ represents the risk threshold equaling disease prevalence (see Webappendix).  MRS is not Net Reclassification Index (NRI)~\citep{Pencina2008}, because under dichotomization, NRI equals Youden's index.

The Appendix shows that MRS measures association by equaling twice the covariance of disease and marker, and is related other association measures.  MRS also equals twice the cross-product difference of the joint probabilities inside a 2x2 table, which is easy to remember by analogy with odds ratios (see Appendix).  

The Webappendix shows the variance of MRS, performance of MRS confidence intervals, and hypothesis testing for two MRSs.

Next we introduce Net Benefit of Information and its relationship to MRS.

\section{Net Benefit of Information and Mean Risk Stratification}
\label{sec:relat-risk-strat}

Decision-theoretic metrics of test performance are based on the expected utility from using the test (see~\cite{Baker2009a} for a comprehensive review).  Calculating expected utility requires specification of the utility for the 4 possible outcomes: the utility of true positive prediction $U_{TP}$, the utility of true negative prediction $U_{TN}$, the utility of false positive prediction $U_{FP}$, and the utility of false negative prediction $U_{FN}$.  Furthermore, the cost of marker $M$ is $U_{Test}$.  These 5 utilities require considering the benefits, harms, and costs of the test and all subsequent interventions, which may be personal and difficult to quantify.    

Instead, rational utility theory requires specification of only a risk-threshold that encapsulates the utilities~\citep{Pauker1980}.  In this approach, the marker/model is dichotomized at risk threshold $R$: $P(D+|M=m_0)=R$.  The $R$ that maximizes expected utility is determined by the ratio of benefit ($B=U_{TP}-U_{FN}$) to costs ($C=U_{TN}-U_{FP}$), which is $R=1/(1+B/C)$~\citep{Pauker1980}.  The risk-threshold weighs the utility of true-positives versus false-positives, e.g. a 10\% threshold means that a rational person is willing to accept 9 false-positives for every true-positive. Decision-theoretic metrics of test performance are plotted versus all possible risk thresholds $R$ (as defined by cutpoints $m_0$) to gain insight on the value of possible risk thresholds.

To derive Net Benefit of Information, we first calculate the expected utility of the test, which averages utilities, weighted by the joint probabilities for the outcomes, plus test cost:
\begin{eqnarray}
U &=& U_{TP}P(D+,M+) + U_{FN}P(D+,M-) + U_{TN}P(D-,M-) + U_{FP}P(D-,M+) + U_{Test}.  \nonumber
\end{eqnarray}
In particular, note that randomly selecting people for intervention, with the same positivity $p=P(M+)$ and cost $U_{Test}$ as the test of interest, has utility~\citep[Ch.9]{Krae:eval:1992}
\begin{eqnarray*}
	U_{RS} = U_{TP}\pi p + U_{FN}\pi(1-p) + U_{TN}(1-p)(1-\pi) + U_{FP}(1-\pi)p + U_{Test},
\end{eqnarray*}
where $\pi=P(D+)$.  Random selection is the minimum utility possible for a test (which need not be zero) and provides a baseline utility that must be substantially exceeded.  A test is more informative the higher its utility is than randomly selecting people for intervention.   

Next, plugging in the following identities (see Webappendix section 2)
\begin{eqnarray*}
	P(D+,M+)&=&MRS/2+p\pi \\ P(D-,M-)&=&MRS/2+(1-p)(1-\pi)\\ P(D+,M-)&=&(1-p)\pi-MRS/2 \\ P(D-,M+)&=&(1-\pi)p-MRS/2
\end{eqnarray*}
into the expected-utility equation yields
\begin{eqnarray*}
	U &=&  \frac{MRS}{2}(U_{TP}-U_{FN}-U_{FP}+U_{TN}) + U_{RS} ~=~ \frac{MRS}{2}(B+C) + U_{RS}. 
\end{eqnarray*}
Finally, we scale utility in units of benefit to define Net Benefit of Information (NBI) as the increase in (scaled) utility from using the test to select people for intervention \textit{versus} randomly selecting people for intervention:
\begin{eqnarray}
NBI(m_0) &=& \frac{U-U_{RS}}{B} = \frac{MRS(m_0)}{2}\frac{B+C}{B} 
= \frac{MRS(m_0)/2}{1-R} \label{NBIdefn}.
\end{eqnarray}
NBI is a function of test characteristics only via the MRS, connecting the decision-theoretic approach to risk-stratification.  For small risk thresholds, NBI is close to half the MRS.  This approximate equivalence provides NBI a concrete risk-stratification interpretation, and \textit{vice versa}, provides a decision-theory justification for MRS.  

Because MRS is a function of the cutpoint $m_0$ that defines the risk threshold $R$ (equation~\ref{eq:1}), so is NBI.  We will plot $NBI(m_0)$ over the range of risk thresholds $R$ defined by cutpoints $m_0$.  The risk thresholds where $NBI(m_0)$ is near its peak is where the most utility is gained over random selection.  This range of risk-thresholds is where the marker/risk-model is "optimally informative".  For small $R$, this range will also maximize $MRS(m_0)$.

\section{Relationship of MRS and NBI to Youden's Index, AUC, and the risk difference}
\label{sec:PropertiesMRSNBI}

\subsection{Relationship of MRS and NBI to Youden's Index and AUC}
\label{sec:YoudenAUC}
MRS and NBI can be calculated by combining prevalence with Youden's index or AUC for a dichotomized marker.  Equation~\ref{eq:MRSdefn} in the Appendix shows that MRS can be written as
\begin{eqnarray*}
	MRS &=& 2\{P(D+,M+)-P(D+)P(M+)\} = 2(Sens-p)\pi, 
\end{eqnarray*}
where sensitivity $Sens=P(M+|D+)$, $\pi=P(D+)$, and $p=P(M+)$.  Denote specificity $Spec=P(M-|D-)$.  Because $p=Sens\times\pi+(1-Spec)(1-\pi)$,
\begin{eqnarray}
MRS(m_0) &=& 2(Sens-Sens\times\pi-(1-Spec)(1-\pi))\pi \nonumber\\ 
&=& 2\pi(1-\pi)J(m_0). \label{eq:2jpi(1-pi)}
\end{eqnarray}
MRS is Youden's index, $J(m_0)=Sens(m_0)+Spec(m_0)-1$, rescaled by disease prevalence.  Note that Youden's index is function of cutpoint $m_0$.  MRS can be calculated by combining an estimate of Youden's index $J$ with an external estimate of disease prevalence $\pi$, which we will do in section~\ref{sec:BRCA_NB_RU}.

Although AUC is usually calculated for continuous markers, for a dichotomized marker, $AUC(m_0)=(J(m_0)+1)/2$~\citep{Cantor2000}.   Thus
\begin{eqnarray}
MRS(m_0) &=& 4(AUC(m_0)-0.5)\pi(1-\pi). \label{eq:4(AUC-0.5)pi(1-pi)}
\end{eqnarray}
Similarly, $NBI(m_0)$ is a function of $J(m_0)$ or $AUC(m_0)$ via the above MRS expressions.  In fact, MRS, NBI, Youden's index, and AUC are maximized when cutpoint $m_0$ implies that risk threshold equals prevalence (see Appendix).  

The key point is that MRS/NBI interpret Youden's index and AUC in light of prevalence.  This is especially important for rare diseases because   
\begin{eqnarray*}
	MRS\approx 2J\pi ~~~\mathrm{and}~~~ NBI\approx J\pi/(1-R).
\end{eqnarray*}
A high Youden's index or AUC might not imply much risk stratification or NBI for rare diseases.  MRS and NBI naturally temper overenthusiasm for markers with high AUC, but for rare diseases.  


\noindent Importantly, disease prevalence bounds MRS and NBI.  For perfect tests ($AUC=1$),
\begin{eqnarray}
max(MRS) &=& 2\pi(1-\pi) \nonumber
\end{eqnarray}
and $NBI=\pi, \forall R$.  Thus if disease is rare, there may be little risk stratification or NBI even for perfect tests.  Figure 1 plots the relationship of MRS to AUC for 3 uncommon disease prevalences. The importance of disease prevalence is illustrated by noting that, the maximum MRS (achieved if AUC=1) is also obtained if AUC=0.55 for diseases 10 times more prevalent; AUC=0.6 suffices if disease is 5 times more prevalent.  Thus a perfect marker for a rare disease provides the same risk-stratification as a weakly-associated marker for a disease that is 5-10 times as prevalent.    

\subsection{Simple and useful decision-theoretic interpretation of Youden's index and AUC}
\label{sec:DecTheoryInterpYoudenAUC}

\begin{figure}[t!]
	\centering
	\includegraphics[angle=-0,width=6in,]{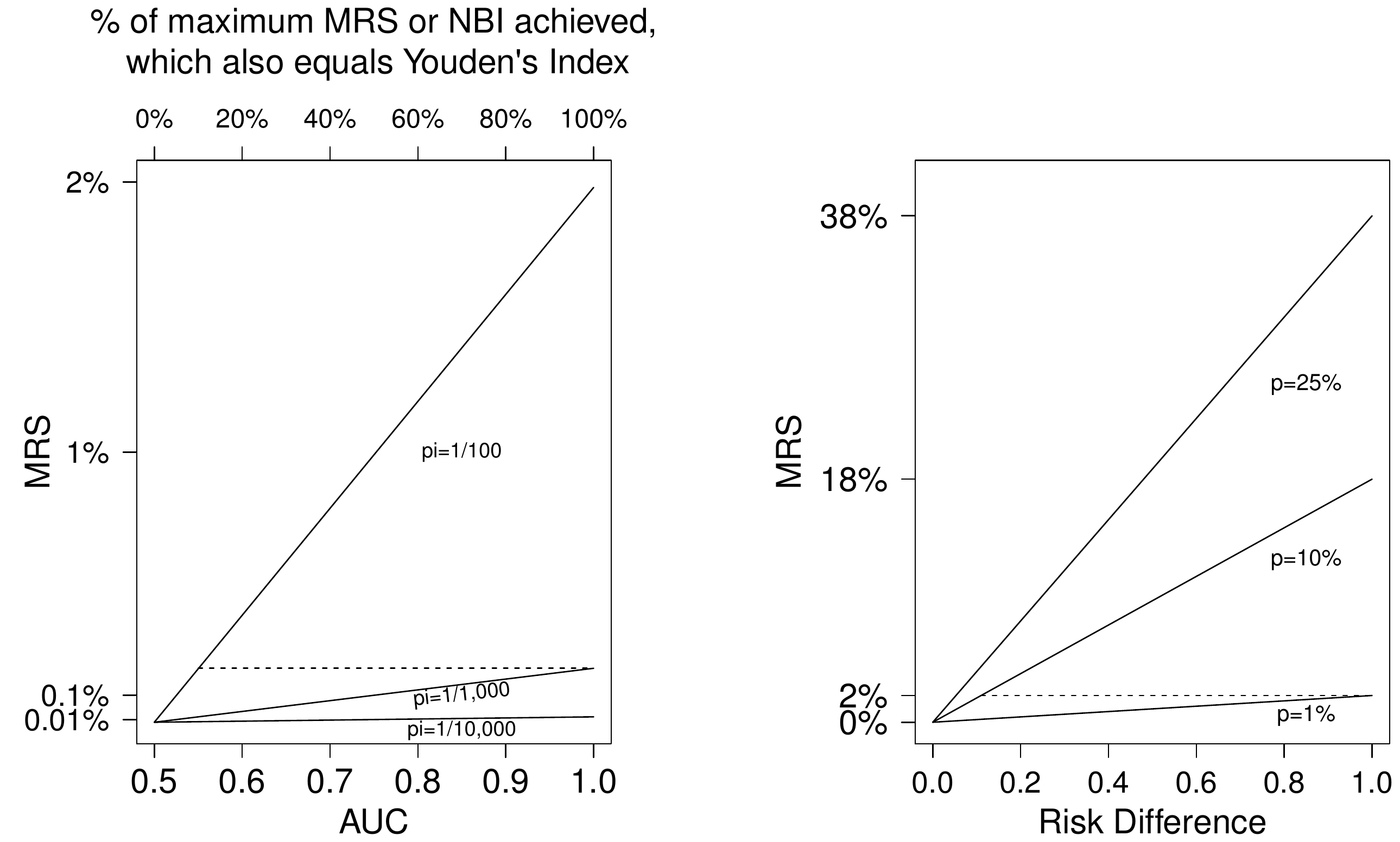}    
	\caption{Left Plot: AUC vs. MRS for 3 disease prevalences (1/100, 1/1,000, and 1/10,000).  Right Plot: Risk-difference vs. MRS for 3 test positivities (1\%, 10\%, and 25\%). Acronyms:  AUC (Area under the receiver operating characteristic curve), MRS (Mean Risk Stratification)}	
	\label{fig:maxmrsnbi}
\end{figure}

The fraction of the maximum NBI and MRS achieved by the test is Youden's index:
\begin{eqnarray*}
	\frac{NBI}{max(NBI)}=\frac{MRS}{max(MRS)} = \frac{2\pi(1-\pi)J(m_0)}{2\pi(1-\pi)} = J(m_0).
\end{eqnarray*}
Youden's index is the fraction of the maximum possible risk-stratification attained by the test.  Youden's index is also the fraction of the maximum possible utility gain over random selection, that is attained by the test.  Thus MRS/NBI indeed provide Youden's index (and thus AUC) with simple and useful decision-theoretic and risk-stratification interpretations. 

But this also illustrates the pitfalls of $J$ and AUC for rare diseases.  Since Youden's index and AUC reflect on multiplicative gains in MRS/NBI, for rare diseases, a high Youden's index or AUC can mask small additive increases in MRS and NBI (Fig.~\ref{fig:maxmrsnbi}, left plot).  Since $J=2\times AUC-1$, a 1\% increase in AUC implies a 2\% increase in MRS or NBI.  Thus MRS/NBI double from AUC=0.6 to 0.7.  An AUC=0.6 is widely considered to be "modest", and indeed, only 20\% of maximal MRS/NBI is achieved.  An AUC=0.7 is widely considered "good", but only 40\% of maximal MRS/NBI is achieved.  An AUC=0.95 is required to achieve 90\% of maximal MRS/NBI.

\subsection{MRS and NBI for a rarely-positive test: relationship to the risk difference}
\label{sec:RarelyPositiveTest}

The risk difference is $t=PPV-cNPV$ (recall $cNPV=P(D+|M-)$).  Risk stratification is sometimes (mis)measured by the risk difference: a large spread in risks is considered evidence of good risk stratification.  Starting from equation~(\ref{eq:MRSdefn}) in the appendix
\begin{eqnarray*}
	MRS &=& 2\{P(D+,M+)-P(D+)P(M+)\} = 2(PPV-\pi)p. 
\end{eqnarray*}
Substituting $\pi = PPV\times p + cNPV(1-p)$,
\begin{eqnarray}
MRS &=& 2(PPV(1-p) - cNPV(1-p))p \nonumber \\
&=& 2tp(1-p). \nonumber
\end{eqnarray}
A large risk difference does not imply much risk stratification, and hence NBI, if the test is rarely positive.  Figure 1 (right panel) plots the relationship of MRS to the risk-difference for 3 test positivity rates. When risk-difference is 1, the maximum MRS is achieved.  The importance of test positivity is illustrated by noting that, the MRS achieved for risk-difference of 1 is also obtained for a risk-difference of approximately only 0.1 when the test is 10 times as positive (dashed line).  Thus a perfect marker for a rarely positive test provides the same risk-stratification, and hence NBI, as a weakly associated marker 10-times as positive.

\section{Informativeness of risk models to select who might get \emph{BRCA1/2} testing}
\label{sec:Example}

As detailed in the Introduction, mutations in the \textit{BRCA1/2} genes cause high risk of breast and ovarian cancers.  The mutations are rare in the general population ($\approx0.25\%$), but 10 times more common among Ashkenazi-Jews.  Currently, women are asked to provide their family history of cancer (e.g. Webappendix Figure 1), and she is offered mutation-testing in the UK and US if a risk model calculates that her risk of carrying a mutation exceeds 10\%~\citep{NICE2017}.  Popular risk models are BRCAPRO~\citep{Parmigiani1998} or BOADICEA~\citep{Antoniou2004}.  We will focus on BRCAPRO.  

However, as mutation-testing costs fall, prominent voices have called for testing all women, which would strain clinical resources by testing millions of women, 99.75\% of whom will test negative.  Instead, a lower risk threshold, below 10\%, might identify nearly all mutation-carriers, yet avoid unnecessary testing for most women.  We recently showed that a low 0.78\% risk-threshold would identify 80\% of Ashkenazi-Jewish mutation-carriers yet test only 44\% of Ashkenazi-Jewish women~\citep{Best2017}.  We use MRS/NBI, AUC and Net Benefit to throw light on the value of the risk model BRCAPRO to select women for \textit{BRCA1/2} testing at risk thresholds between 0\%-10\%.

We use data on 4,589 volunteers (102 \textit{BRCA1/2} mutation carriers) from the Washington Ashkenazi Study (WAS)~\citep{STRUEWING1997}.  We calculated each volunteer's risk of carrying a mutation, based on their self-reported family-history of breast/ovarian cancers, using BRCAPRO.  Here $M$ is the BRCAPRO risk score, and because BRCAPRO is a well-calibrated risk model~\citep{Best2017}, $m_0=R$, i.e. the cutpoint $m_0$ equals the risk threshold $R$. Disease $D$ indicates the presence of a \textit{BRCA1/2} mutation.

\subsection{MRS and NBI for BRCAPRO at different risk-thresholds for Ashkenazi-Jews}
\label{sec:MRS_NBI_AJs}

\begin{figure}[t!]
	\includegraphics[angle=-0,width=7.4in,]{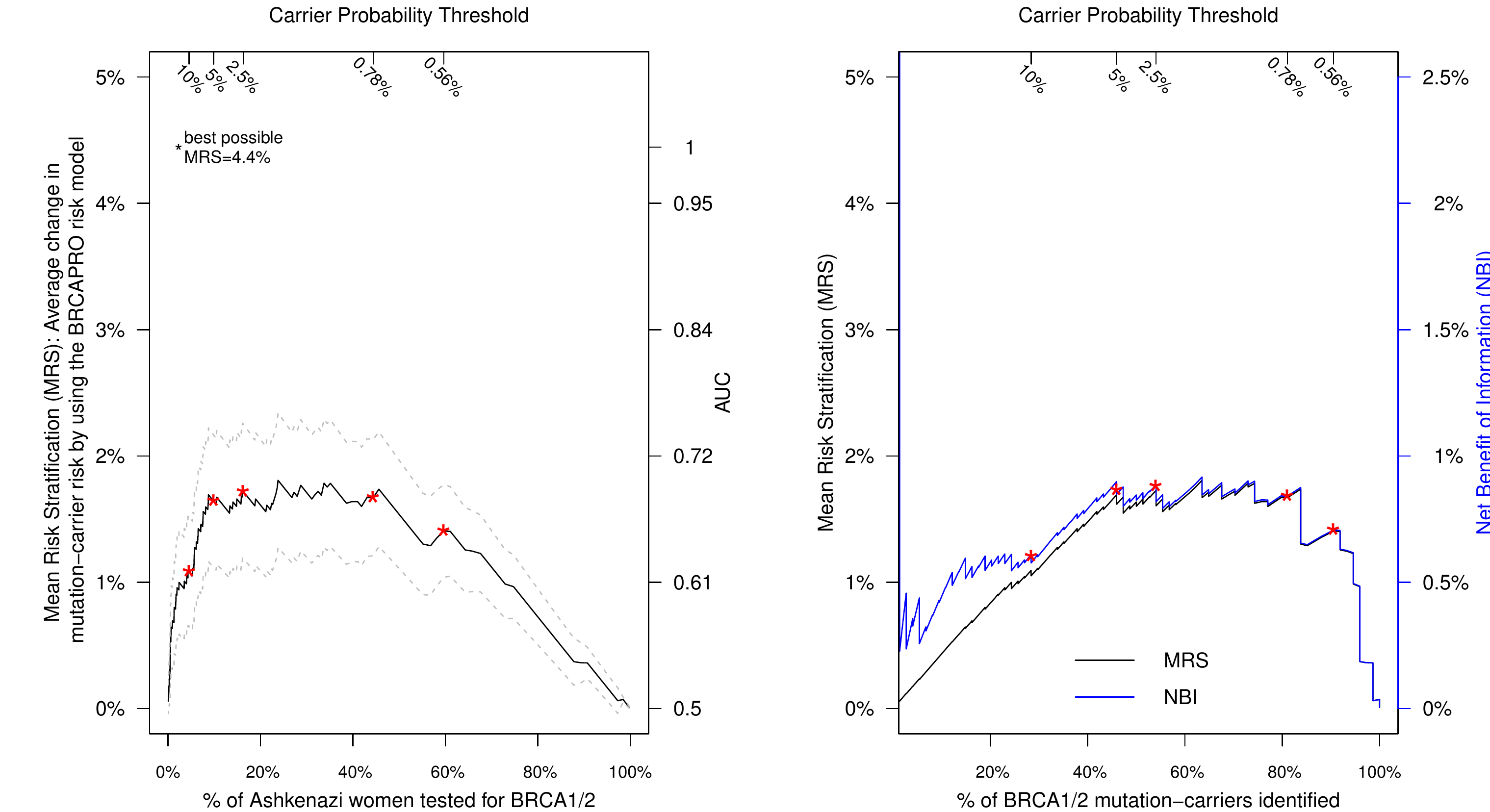}
	\caption{The left plot is MRS vs. \% of Ashkenazi-Jews who would be referred for testing using BRCAPRO at various risk thresholds (top axis).  The right axis is the AUC corresponding to MRS.  The gray lines are 95\% pointwise confidence bands for MRS (see Appendix for variance of MRS).  The right plots MRS and Net Benefit of Information (NBI) vs. sensitivity (\% of \textit{BRCA1/2} mutations detected) for the BRCAPRO risk model to detect \textit{BRCA1/2} mutation-carriers.  Note that the x-axes differ in the two plots.}
	\label{fig:BRCANNTestMRS}
\end{figure}

Recall that $MRS(m_0)$ is a function of the cutpoint used to decide $M+$ (equation~\ref{eq:1}).  The left panel of Figure~\ref{fig:BRCANNTestMRS} plots $MRS(m_0)$ for BRCAPRO over a range of risk thresholds $R$ ($=m_0$) (top axis) for Ashkenazi-Jews.  The x-axis is the fraction of women testing positive by each risk threshold. The right axis is the AUC implied by each MRS, according to equation~(\ref{eq:4(AUC-0.5)pi(1-pi)}), given the observed $\pi=2.3\%$ mutation prevalence for Ashkenazi-Jews. 

BRCAPRO has best $MRS\approx1.7\%$ for risk thresholds in a "sweetspot" of 0.78\% to 5\%.  An MRS=1.7\% means that a woman who uses BRCAPRO, dichotomized at any threshold between 0.78\% to 5\% to refer for mutation-testing, will learn that her risk of carrying a mutation will increase or decrease by 1.7\% on average.  An average change in risk of $\pm1.7\%$ seems meaningful because it is similar to pretest risk (i.e. mutation prevalence) of 2.3\%.   

In contrast, the current 10\% threshold yields a substantially lower MRS of $1.1\%$ (p=0.039 versus MRS=1.7\% at the 0.78\% threshold).  The Webappendix shows how to calculate the variance of MRS and conduct hypothesis testing for two MRSs.  

The 10\% threshold yields a much higher risk-difference than the 0.78\% threshold: 12.58\% vs. 3.39\% respectively.  Thus MRS was lower at the 10\% threshold, in spite of a higher risk-difference, because the 10\% threshold has only 4.5\% test-positivity (vs. 44\% at the 0.78\% threshold).  Rarely positive tests have low risk-stratification (see section~\ref{sec:RarelyPositiveTest}).

The right-axis of figure~\ref{fig:BRCANNTestMRS} (left panel) shows the AUC implied by each risk threshold.  The 0.78\%-5\% risk-threshold sweetspot has $AUC\approx0.69$, indicating that only 38\% of the maximum MRS of 4.4\% is achievable by BRCAPRO.  The MRS=1.7\% reveals the risk-stratification implications of AUC=0.69.  

$MRS(m_0)$ is maximized when the risk threshold equals disease prevalence, i.e. $m_0=R=\pi$ (see Appendix).  Thus the sweetspot of risk thresholds that maximizes MRS/NBI, and Youden's index and AUC, always includes prevalence, in this case, 2.3\%.

Recall that, just as MRS is a function of the risk-threshold cutpoint $m_0$, so is NBI (equation~\ref{NBIdefn}).  Figure~\ref{fig:BRCANNTestMRS} (right panel) examines $MRS(m_0)$ and $NBI(m_0)$ versus the sensitivity (\% of \textit{BRCA1/2} mutations detected) of BRCAPRO as the risk threshold varies. This plot trades off MRS/NBI, which informs a woman of the informativeness of BRCAPRO for herself, versus sensitivity, which is the public-health perspective of identifying as many mutation-carriers as possible.  The MRS/NBI risk-threshold sweetspot of 0.78\%-5\% identifies 45\%-80\% of \textit{BRCA1/2} mutation-carriers, while the traditional 10\% threshold identifies only 28\% of \textit{BRCA1/2} mutation-carriers.  In the 0.78\%-5\% risk-threshold range, the NBI of around 0.85\% means that BRCAPRO additively increases utility by 0.85\% over random selection.  More importantly, MRS and NBI/2 are very similar for the low risk thresholds of interest (below 10\%).  Thus the ranges of risk thresholds chosen by MRS and NBI will coincide and maximize both risk stratification and utility gain over random selection.  Hence, BRCAPRO is optimally informative, when dichotomized at risk-thresholds between 0.78\%-5\%, to identify which Ashkenazi-Jewish women to refer for \textit{BRCA1/2} testing.

\subsection{MRS/NBI and Net Benefit: Complementary perspectives}
\label{sec:MRSvsNB}

Net Benefit is an important modern approach to identify risk-thresholds where a marker/model is useful for clinical actions~\citep{Vickers2006}.  We summarize Net Benefit, demonstrate its relationship to NBI, and compare insights from Net Benefit and MRS/NBI on risk thresholds for \textit{BRCA1/2} testing. 

Recall that NBI subtracts the utility of random selection $U_{RS}$ from the utility of the test $U$, standardized by benefit $B$ (see section~\ref{sec:relat-risk-strat}).  In contrast, Net Benefit (NB) subtracts the utility of calling everyone negative ($U_N=U_{FN}\pi+U_{TN}(1-\pi)$) from the utility of the test, standardized by benefit:
\begin{eqnarray*}
	NB = \frac{U-U_N}{B} &=& \pi Sens-\frac{R}{1-R}(1-Spec)(1-\pi).
\end{eqnarray*}
In particular, the Net Benefit of calling everyone positive subtracts the the utility of calling everyone negative from the utility of calling everyone positive $U_P = U_{FP}(1-\pi)+U_{TP}\pi$:
\begin{eqnarray*}
	NB_P = \frac{U_P-U_N}{B} &=& \pi - \frac{R}{1-R}(1-\pi).
\end{eqnarray*}
The goal of Net Benefit is to identify the range of risk thresholds where the marker/model provides more utility than all-or-nothing actions.  These thresholds will be those where the Net Benefit is positive (where the test provides more utility than calling everyone negative) and greater than $NB_P$ (where the test provides more utility than calling everyone positive).  

Note that the Net Benefit of random selection (i.e. $Sens=1-Spec$) is not zero:
\begin{eqnarray*}
	NB_{RS} = \frac{U_{RS}-U_N}{B} &=& \pi Sens - \frac{R}{1-R}Sens(1-\pi). 
\end{eqnarray*}
NBI equals the Net Benefit of the test minus the Net Benefit of random selection:
\begin{eqnarray*}
	NBI = \frac{U-U_N - (U_{RS}-U_N)}{B} &=& NB - NB_{RS}.
\end{eqnarray*}

\begin{figure}[t!]
	\includegraphics[angle=-0,width=7in,]{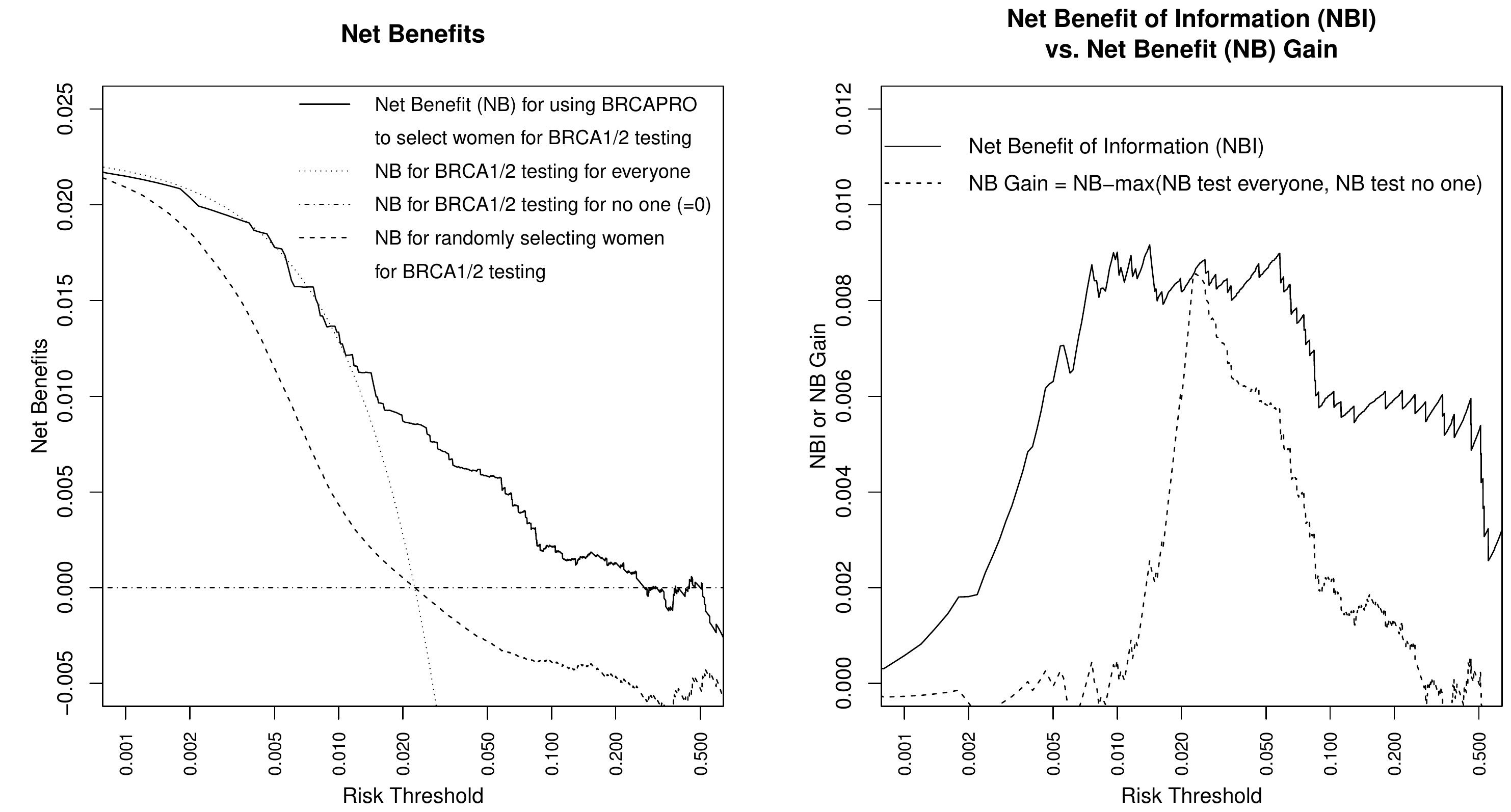}
	\caption{The left plot is Net Benefit for the BRCAPRO risk model at different risk thresholds for Ashkenazi-Jews.  The right plot compares Net Benefit of Information for BRCAPRO to the increase in Net Benefit vs testing all Ashkenazi-Jews or none (Net Benefit Gain), by using BRCAPRO.}
	\label{fig:NBvsNBI}
\end{figure}

The left panel of Figure~\ref{fig:NBvsNBI} shows the Net Benefit for using the BRCAPRO risk model (solid line) to the Net Benefit for doing \textit{BRCA1/2} testing on everyone ($NB_P$; dotted) and the Net Benefit for \textit{BRCA1/2} testing for no one (zero by definition).  The Net Benefit of randomly selecting women for \textit{BRCA1/2} testing (dashed line) is always between the Net Benefits for testing everyone or no one (all 3 are zero at risk threshold equals prevalence).  

Note that Net Benefit is greater than $NB_P$ starting at risk thresholds around 1.7\%, and remains positive until around 30\%.  According to Net Benefit, the BRCAPRO risk model appears useful for risk thresholds between 1.7\%-30\%.  Below 1.7\%, all Ashkenazi-Jews should be referred for mutation testing, and above 30\%, none should be referred.  Although 1.7\% is within the 0.78\%-5\% MRS "sweetspot", the BRCAPRO model is rather uninformative at risk thresholds above 10\%.  For example, the 30\% threshold has MRS=0.79\% (p=0.0005 vs. 1.7\%) and AUC=0.59.  For a 2.3\% prevalence, an average change of $\pm0.79\%$ seems small.

Figure~\ref{fig:NBvsNBI} (right panel) compares NBI to Net Benefit Gain, $NB-max(NB_P,0)$: the increase in Net Benefit versus testing everyone or no one. NBI and Net Benefit coincide at risk threshold equals prevalence (see Webappendix), where they and MRS achieve their maximum (see Appendix).  Thus the ranges of useful risk-thresholds, as determined by Youden's index, AUC, MRS/NBI and Net Benefit, will all overlap at disease prevalence.

Note that Net Benefit Gain is nearly zero for risk thresholds 0.5\%-1\%, where NBI remains near its peak.  MRS/NBI and Net Benefit answer different questions and thus present different perspectives.  At the 0.78\% risk-threshold, MRS/NBI are at their peak and the BRCAPRO risk-model is optimally informative, but Net Benefit implies that the model should not be used and all Ashkenazi-Jews should undergo \textit{BRCA1/2} testing.  MRS/NBI emphasize that the model is optimally informative at 0.78\%, identifying 80\% of \textit{BRCA1/2} mutation-carriers while testing only 44\% of Ashkenazi-Jews.  In contrast, Net Benefit notes that a 0.78\% threshold implies that a rational person trades-off 127 false-positives for 1 true-positive.  False-positives are unimportant, and one should not use the model (even though it is optimally informative) but rather refer all Ashkenazi-Jews for \textit{BRCA1/2} testing. 

In summary, Net Benefit emphasizes that some risk thresholds are so low that false-positives are unimportant and thus everyone should be tested.  MRS/NBI emphasize that the risk model is optimally informative, refers only a minority for \textit{BRCA1/2} testing yet identifies the big majority of mutation-carriers.  We find both perspectives illuminating.

\subsection{Comparing MRS/NBI to other risk prediction metrics: Importance of prevalence}
\label{sec:BRCA_NB_RU}

\begin{figure}[t!]
	\centering\includegraphics[angle=-0,width=7in,]{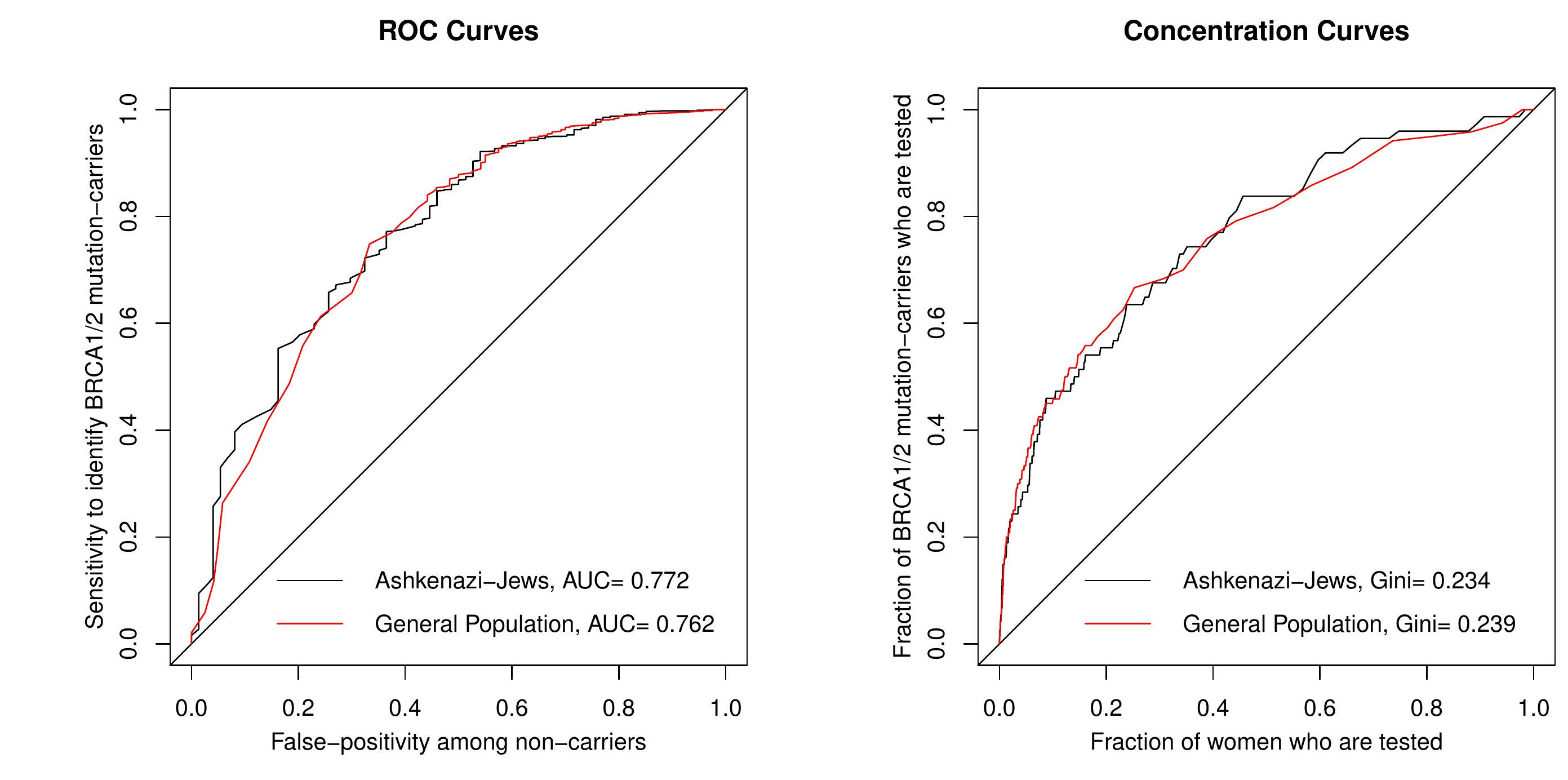}    
	\caption{ROC curves (left plot) and Concentration (Lorenz) curves (right plot) for the BRCAPRO risk model to predict if a woman carries a \textit{BRCA1/2} mutation, for Ashkenazi-Jews and the general population.  Note the curves are similar between populations, even though mutation prevalence is 10-times higher among Ashkenazi-Jews.}
	\label{fig:LorenzCurve}
\end{figure}

\textit{BRCA1/2} mutations induce the same cancer risk for Ashkenazi-Jews and the general-population~\citep{Kuchenbaecker2017}.  Only mutation prevalence varies between populations.  This situation allows us to isolate the effect of prevalence on risk-prediction metrics.

Because we do not have comparable data on \textit{BRCA1/2} mutations in the general-population, we  will approximate MRS/NBI for the general-population by combining the general-population mutation-prevalence with sensitivity/specificity from the WAS (see section~\ref{sec:YoudenAUC}).  Because pathogenic \textit{BRCA1/2} mutations have the same risk for cancer regardless of population, the Bayes Factor used by BRCAPRO is the same regardless of population; only the prior distribution (i.e. mutation prevalence) differs by population~\citep{Katki2005}.   Thus we use BRCAPRO to calculate mutation carrier-risks on WAS data, but substitute the mutation prevalence in the general-population (0.26\%) as the prior.  To calculate specificity and prevalence, we weight non-carriers by a factor of 9 to ensure that mutation-prevalence is 10-times smaller than in the WAS.  Figure~\ref{fig:LorenzCurve} shows that Ashkenazi-Jews and the general population indeed have similar ROC curves and Lorenz curves.  This reinforces that ROC, AUC, and Lorenz curves cannot distinguish between populations when only disease prevalence differs between them.

\begin{figure}[t!]
	\centering\includegraphics[angle=-0,width=7in,]{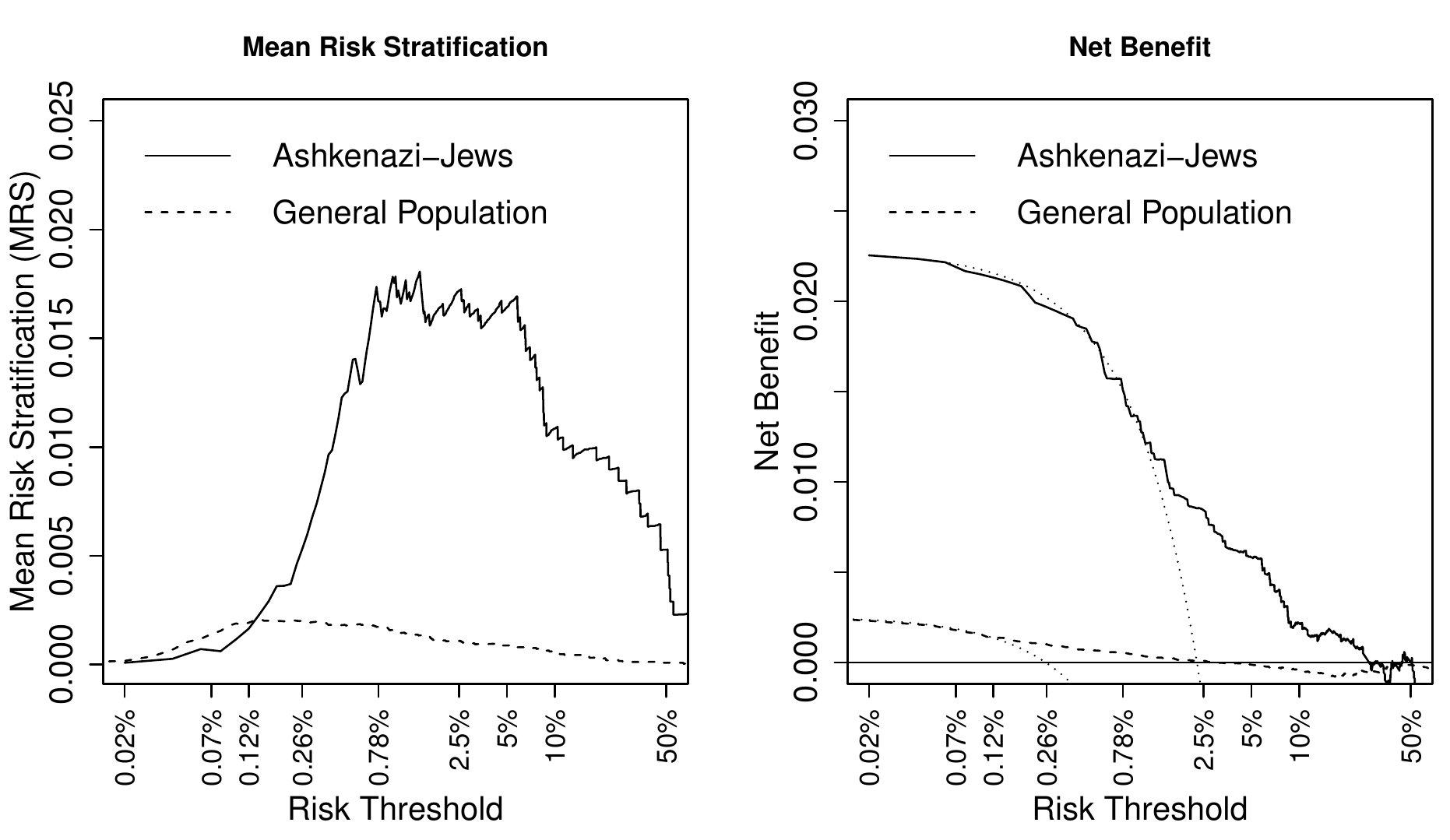}
	\caption{Performance of Mean Risk Stratification and Net Benefit for using BRCAPRO to decide who should get \textit{BRCA1/2} mutation testing, among Ashkenazi-Jews (solid lines) and the general population (dashed lines).  The dotted lines in the Net Benefit plot are the Net Benefits if everyone in each population gets \textit{BRCA1/2} testing.}
	\label{fig:BRCA_NB_RU_AJgenpop}
\end{figure}

Figure~\ref{fig:BRCA_NB_RU_AJgenpop} examines MRS and Net Benefit among Ashkenazi-Jews and in the general population.  Because NBI and MRS/2 in the general population are almost identical for risk thresholds below 10\% (not shown), we focus on MRS. For the general population, the MRS/NBI sweetspot (left panel) of 0.07\%-0.56\% yields MRS of only around 0.20\%.  In this sweetspot, the AUC=0.69, the same as for Ashkenazi-Jews in their sweetspot of 0.78\%-5\%.  However, the MRSs are very different because prevalences are very different.  MRS and NBI reveal the very different risk-stratification, and decision-theoretic, implications of equal AUC in populations with different prevalences.  

Both MRS/NBI and Net Benefit (Fig.~\ref{fig:BRCA_NB_RU_AJgenpop}, right panel) concur that BRCAPRO is generally more useful for Ashkenazi-Jews than the general-population.  However, MRS notes that, if the risk threshold were below 0.12\%, then BRCAPRO is actually more informative in the general population.  Below 0.12\% there are hardly any mutation-carriers among Ashkenazi-Jews, but 33\% of mutation-carriers remain in the general population.

According to Net Benefit, the BRCAPRO model in the general-population is best between risk thresholds of 0.12\% to 2.5\%.  In this range, AUC drops from 0.69 to 0.61, and MRS is nearly halved from 0.20\% to 0.11\%.  Given that mutation-prevalence is 0.26\%, reductions in MRS much below 0.20\% seem like substantial losses of informativeness.

At certain low risk thresholds, such as 0.26\%, Net Benefit suggests mutation-testing for all Ashkenazi-Jews but using the BRCAPRO model to select general-population women for mutation testing.  MRS/NBI can be seen to agree with Net Benefit.  For Ashkenazi-Jews, the model has rather low MRS at 0.26\% compared to other risk thresholds, suggesting that the BRCAPRO model is not useful for Ashkenazi-Jews at a 0.26\% threshold.  For the general-population, the model is optimally informative at 0.26\%, although both the MRS and Net Benefit are low. 

Dismayingly, at the current 10\% risk-threshold in the general-population, Net Benefit finds that \textit{no one} should undergo \textit{BRCA1/2} testing (i.e. the dashed line is below zero at 10\%).  In contrast, MRS/NBI note only that the risk-model is uninformative at the 10\% threshold (MRS=0.05\%), and makes no judgment about the value of \textit{BRCA1/2} testing.

\section{Discussion}
\label{sec:discuss}

To better understand the diagnostic performance of markers and risk-models at various risk thresholds, we introduced two new broadly applicable, linked metrics: Mean Risk Stratification (MRS) and Net Benefit of Information (NBI).  We presented 4 key results.  First, Youden's index and AUC reflect on (1) the fraction of the maximum possible utility gain over random selection (i.e. NBI) that is attained by the test and (2) the fraction of the maximum possible risk-stratification (i.e. MRS) that is attained by the test.  MRS and NBI provide Youden's index and AUC with decision-theoretic and risk-stratification rationale, the lacks of which have long been a criticism of Youden's index and AUC.  Second, for rare diseases, high AUC does not imply high risk-stratification or utility gain over random selection.  AUC must be considered in light of disease prevalence, which is automatically done by MRS and NBI.  Third, NBI is a function of only MRS and the risk-threshold for action, connecting decision-theory to risk-stratification and providing a decision-theoretic rationale for MRS.  Last, MRS and NBI provide a range of risk thresholds for which the risk model is "optimally informative": risk-thresholds that maximize both risk-stratification and the utility gain over random selection.

We proposed an eclectic approach, using AUC, MRS/NBI, and Net Benefit, to evaluate risk-thresholds for the BRCAPRO risk model to refer women for \textit{BRCA1/2} mutation-testing.  Although the AUC is essentially the same for both the general population and for Ashkenazi-Jews, both MRS/NBI and Net Benefit note that mutation testing is more valuable for Ashkenazi-Jews, for whom mutations are 10-times more prevalent.  Interestingly, MRS/NBI notes that at extremely low risk thresholds, BRCAPRO is actually more informative in the general population, because these risk-thresholds are so low that nearly all Ashkenazi-Jews would be tested anyway.  

MRS/NBI and Net Benefit address complementary questions: MRS/NBI quantify the utility of the information in a test, while Net Benefit quantifies the utility versus all-or-nothing actions.  In the \textit{BRCA1/2} example, both perspectives were valuable.  Net Benefit emphasizes that risk thresholds below 1.7\% are so low that false-positives are unimportant and thus all Ashkenazi-Jews should be tested; instead, risk-thresholds up to 30\% could be considered for using the BRCAPRO model.  MRS/NBI emphasize that for risk thresholds in 0.78\% to 5\%, the BRCAPRO model is optimally informative, referring only a minority of Ashkenazi-Jews for \textit{BRCA1/2} testing yet identifying the big majority of mutation-carriers.  At these risk thresholds, the MRS=1.7\%, meaning that a woman who uses BRCAPRO will learn that her risk of carrying a mutation will increase or decrease by 1.7\% on average.  An average change in risk of 1.7\% seems meaningful because it is similar to pretest risk (i.e. mutation prevalence) of 2.3\%.  MRS/NBI note a substantial loss of BRCAPRO model informativeness at risk thresholds above 5\%, and especially 30\%.  The ranges of useful risk-thresholds, as determined by Youden's index, AUC, MRS, NBI, and Net Benefit, will always overlap at risk threshold equaling disease prevalence.

Dismayingly, at the current 10\% risk-threshold in the general-population, Net Benefit finds that \textit{no one} should undergo \textit{BRCA1/2} testing.  According to Net Benefit, \textit{BRCA1/2} testing is never worthwhile at a 10\% threshold, because it implies that one is willing to trade-off only 9 false-positives for 1 true-positive.  However, current \textit{BRCA1/2} testing at 10\% is widely agreed to be a major success~\citep{GenomeWeb2017} and is recommended in the US~\citep{Moyer2014} and by the UK NICE~\citep{NICE2017}.  Stopping all \textit{BRCA1/2} testing would be considered absurd.  This example suggests limits for applying rational risk-threshold theory to develop medical guidelines.  In contrast, MRS/NBI make no judgment about the value of the actual intervention (i.e. genetic testing), only noting that the BRCAPRO model at the 10\% threshold is rather uninformative.  

However, we caution that neither MRS/NBI nor Net Benefit can determine the risk threshold for action.  Prespecified utilities determine risk-thresholds.  If a threshold is not optimally informative, that fact should be noted but does not disqualify the threshold.

MRS/NBI reinforce that disease prevalence and test-positivity are crucial for evaluating risk-stratification and interpreting AUC.  An AUC=0.6 achieves only 20\% of maximum risk-stratification/NBI and AUC=0.95 is required to achieve 90\%.  Furthermore, there is little risk-stratification or NBI possible for rare diseases or for rarely positive tests.  MRS/NBI readily interpret AUC in light of disease prevalence, making it immediately clear why \textit{BRCA1/2} testing is less valuable in the general population, where mutations are rare, but might be more valuable for Ashkenazi-Jews, where mutations are more common.  Although experts are aware of the importance of disease prevalence~\citep[Sec. 10]{Baker2009a}, or can fix the ROC curve to account for prevalence~\citep{Hilden1991} (see Webappendix), MRS is a simple metric, intuitive to scientists, that automatically does the job.  MRS/NBI could be routinely calculated to better interpret AUC.

Although we introduced both MRS and NBI, we prefer to focus on MRS when possible.  MRS has a concrete risk-stratification interpretation, which we find to be more appealing to scientists than the abstract "utility" interpretation of decision-theoretic metrics.  MRS measures association via its simple formula as twice the cross-product difference of joint probabilities in a 2x2 table, which is easy to grasp in analogy with odds ratios.  At low risk thresholds, as would be reasonable for rare conditions such as cancer or \textit{BRCA1/2} mutations, the NBI is close to MRS/2 and thus either NBI or MRS could be used.  The main value of NBI is as a decision-theoretic rationale for using MRS in practice.

MRS/NBI are valuable to include in an eclectic approach to evaluating tests that includes AUC and Net Benefit.  MRS/NBI reveals the risk-stratification meaning of AUC, and provides a complementary perspective to Net Benefit for considering risk thresholds for action.  Much work remains to be done to make MRS/NBI usable in practice, especially for longitudinal markers and models necessary for disease screening~\citep{Sweeting2012}.  
\href{http://analysistools.nci.nih.gov/biomarkerTools}{Our MRS webtool is available.}

\section*{Acknowledgements}

This research was supported by the Intramural Research Program of the NIH/NCI.  We thank Mark Schiffman and Anil Chaturvedi for their long-standing support and discussions.  We thank Ionut Bebu and Holly Janes for valuable comments on prior drafts.  We are indebted to our late mentor, collaborator, and friend Sholom Wacholder for his support.  We thank Christine Fermo and Sue Pan for helping develop the MRS Webtool. 
\vspace*{-8pt}

\appendix

\section*{Appendix}
\label{Appendix}

\subsection*{MRS is maximized when dichotomizing at disease prevalence}
\label{maxMRSNBIatPrevalence}

Equation~\ref{eq:2jpi(1-pi)} notes that $MRS(m_0)=2J(m_0)\pi(1-\pi)$, where $J(m_0)$ is Youden's index calculated at cutpoint $m_0$.  MRS is maximized as a function of cutpoint $m_0$ when Youden's index $J(m_0)$ is maximized, which occurs when dichotomizing at disease prevalence: $P(D+|M=m_0)=P(D+)=\pi$.  To prove this we differentiate Youden's index as a function of $m_0$ 
\begin{eqnarray*}
	J(m_0) &=& \int_{m_0}^\infty P(M=m|D+)-P(M=m|D-)~dm
\end{eqnarray*}
with respect to the cutpoint $m_0$ using the Leibnitz Integral Rule:
\begin{eqnarray*}
	\frac{dJ(m_0)}{dm_0} &=& P(M=m_0|D-)-P(M=m_0|D+).
\end{eqnarray*}
Setting the derivative equal to zero, and using Bayes' rule:
\begin{eqnarray*}
	\frac{P(D+|M=m_0)P(M=m_0)}{\pi} &=& \frac{P(D-|M=m_0)P(M=m_0)}{1-\pi}\\
	P(D+|M=m_0) &=& \pi
\end{eqnarray*}
Thus dichotomizing marker/model $M$ at $m_0$ at disease prevalence $\pi$ maximizes Youden's index and thus MRS.  Thus the "sweetspot" of risk-thresholds that maximize MRS will always include disease prevalence.  At this cutpoint, MRS equals Total Gain~\citep{bura2001binary} (see Webappendix).

\subsection*{MRS measures association: MRS is twice the covariance of disease and marker}
\label{sec:Association}

Recall that $PPV=P(D+|M+)$ and $cNPV=P(D+|M-)$.  Rewriting MRS equation~(\ref{eq:1}): 
\begin{eqnarray}
MRS &=& \{PPV-P(D+)\}P(M+) + \{P(D+)-cNPV\}P(M-) \nonumber\\
&=& \{PPV-P(D+)\}P(M+) + P(D+)\{1-P(M+)\} - \{P(D+)-P(D+,M+)\} \nonumber\\
&=& 2\{P(D+,M+) - P(D+)P(M+)\}.   \label{eq:MRSdefn}
\end{eqnarray}
MRS is simply twice the covariance of $D$ and $M$.  MRS is zero if and only if disease and marker are independent.  Negative MRS means that $M+$ is inversely associated with disease.  When test positive/negative are interchanged, MRS changes sign.  

Other association measures, such as Pearson's correlation, the Phi coefficient, Yule's Q, and Cohen's Kappa, use MRS as a numerator but standardize it with different denominators.  MRS is the numerator of the Mantel-Haenszel and Cochran's tests~\citep[Ch 2.6]{Lac00}.  

The Webappendix demonstrates that MRS also equals the departure of \textit{any} of the 4 joint probabilities of $D$ and $M$ from the product of their margins:
\begin{eqnarray*}
	MRS &=& 2\{P(D+,M+) - P(D+)P(M+)\} = 2\{P(D+)P(M-) - P(D+,M-)\} \\
	&=& 2\{P(D-,M-) - P(D-)P(M-)\} = 2\{P(D-)P(M+) - P(D-,M+)\}\label{eq:4MRS}
\end{eqnarray*}

\subsection*{MRS is twice the cross-product difference of joint probabilities inside 2x2 tables}
\label{sec:CrossProductDiff}

Denote $a=P(D+,M+),~b=P(D+,M-),~c=P(D-,M+),~d=P(D-,M-)$.  Substituting into MRS equation~(\ref{eq:MRSdefn}) (see above) yields
\begin{eqnarray*}
	MRS &=& 2\{P(D+,M+) - P(D+)P(M+)\} = 2\{a - (a+b)(a+c)\} \\
	&=& 2\{a(1-a) - ab - ac - bc\}\\
	&=& 2\{a(b+c+d) - ab - ac - bc\}\\
	&=& 2\{ad-bc\} \label{eq:19} 
\end{eqnarray*}
MRS is simply twice the cross-product \textit{difference} of the joint probabilities in the interior of the 2x2 table.  The cross-product difference is also the determinant of the 2x2 table as a matrix.  In contrast, the odds ratio (OR) is the cross-product \textit{ratio}.  Being a ratio, the OR is dimensionless, while the MRS is on the scale of risk differences.  MRS as a cross-product difference is easy for scientists to remember.

\section*{Web Appendix}
Additional materials can be found in the file of WebAppendix at the end.



\newpage
\setcounter{page}{0}   
\pagenumbering{roman}  

\begin{center}
\Large\emph{\textbf{WebAppendix for the Paper: }}

\vspace{3mm}
\textbf{``Novel decision-theoretic and risk-stratification metrics of predictive performance: Application to deciding who should undergo genetic testing"}

\vspace{3mm}
\emph{by} Hormuzd A. Katki
\end{center}




\section{Total Gain is a special case of Mean Risk Stratification}
\label{TotalGain}
Total Gain (TG) measures the explanatory power of a continuous covariate $M$ in a binary regression model $P(D+|M=m)=\pi(m)$, where $\pi(m)$ is typically a logistic regression~\citep{bura2001binary}.  Denote overall disease prevalence as $P(D+)=\pi$.  Choose cutpoint $m^*$ such that $M$ is cut at disease prevalence: $P(D+|M=m^*)=\pi(m^*)=\pi$.  Then
\begin{eqnarray*}
	TG &\equiv& 2\left| \int_{m^*}^{\infty} (\pi(m)-\pi)~dF(m) \right| = 2|P(D+,M\ge m^*) - P(D+)P(M\ge m^*)|.
\end{eqnarray*}
Thus TG is $|MRS|$ for continuous $M$ cut at disease prevalence.  MRS is valid for discrete/continuous $M$ and allows any cutpoint of $M$.  Unlike MRS, TG is non-negative.

\section{MRS equals twice the deviation of any joint probability from the product of its marginals}
\label{sec:4MRSeqns}

This section proves that MRS can be written as twice the deviation from any of the 4 joint probabilities inside a 2x2 tables from the product of its corresponding marginals.  Recall that $PPV=P(D+|M+)$ and $cNPV=1-NPV=P(D+|M-)$.  Thus 
\begin{eqnarray*}
	MRS &=& \{PPV-P(D+)\}P(M+) + \{P(D+)-cNPV\}P(M-) \\
	&=& \{P(D+,M+))-P(D+)P(M+)\} + \{P(D+)P(M-)-P(D+,M-)\}.
\end{eqnarray*}
Denote $P(D+)=\pi$ and $P(M+)=p$.  First, substituting $P(D+,M-)=\pi-P(D+,M+)$ yields MRS as a deviation from $P(D+,M+)$:
\begin{eqnarray*}
	MRS &=& \{P(D+,M+))-p\pi\} + \pi\{1-p\} - \{\pi-P(D+,M+)\} \\
	&=& \{P(D+,M+))-p\pi\} - p\pi + P(D+,M+) \\
	&=& 2\{P(D+,M+) - p\pi\}.
\end{eqnarray*}
Thus $P(D+,M+)=MRS/2 + p\pi$.  

Second, substituting $P(D+,M+)=\pi-P(D+,M-)$ yields MRS as a deviation from $P(D+,M-)$:
\begin{eqnarray*}
	MRS &=& 2\{P(D+,M+) - p\pi\} \\
	&=& 2\{\pi-P(D+,M-) - p\pi\} \\
	&=& 2\{\pi(1-p) - P(D+,M-)\}.
\end{eqnarray*}
Thus $P(D+,M-)=\pi(1-p)-MRS/2$.

Third, substituting $P(D+,M-)=(1-p)-P(D-,M-)$ yields MRS as a deviation from $P(D-,M-)$:
\begin{eqnarray*}
	MRS &=& 2\{\pi(1-p) - P(D+,M-)\} \\
	&=& 2\{\pi(1-p) - \{(1-p)-P(D-,M-)\}\} \\
	&=& 2\{(1-p)\{\pi-1\} +P(D-,M-)\} \\
	&=& 2\{P(D-,M-) - (1-\pi)(1-p)\}.
\end{eqnarray*}
Thus $P(D-,M-) = MRS/2 + (1-p)(1-\pi)$.

Fourth and final, substituting $P(D-,M-)=(1-\pi)-P(D-,M+)$ yields MRS as a deviation from $P(D-,M+)$:
\begin{eqnarray*}
	MRS &=& 2\{P(D-,M-) - (1-\pi)(1-p)\} \\
	&=& 2\{(1-\pi)-P(D-,M+) - (1-\pi)(1-p)\}\\
	&=& 2\{(1-\pi)(1-(1-p)) - P(D-,M+)\}\} \\
	&=& 2\{p(1-\pi) - P(D-,M+)\}.
\end{eqnarray*}
Thus $P(D-,M+) = p(1-\pi) - MRS/2$.

\section{Relationship of Net Benefit to Net Benefit of Information}
\label{(NBandNBI}

Denote $P(D+)=\pi$ and $P(M+)=p$.  Net Benefit (NB), ignoring test costs, is
\begin{eqnarray*}
	NB &=& \pi Sens-\frac{R(1-Spec)(1-\pi)}{1-R} \\
	&=& P(D+,M+)-\frac{R}{1-R}P(D-,M+) \\
	&=& \frac{P(D+,M+)-pR}{1-R},
\end{eqnarray*}
by substituting $P(D-,M+)=P(M+)-P(D+,M+)$.  Then
\begin{eqnarray*}
	NB &=& \frac{PPV-R}{(1-R)/p}.
\end{eqnarray*}
Now starting from NBI, and substituting MRS equation (7) from the main paper
\begin{eqnarray*}
	NBI &=& \frac{MRS/2}{1-R}\\
	&=& \frac{P(D+,M+)-P(D+)P(M+)}{1-R}\\
	&=& \frac{P(D+,M+)-p\pi}{1-R} \\
	&=& \frac{PPV-\pi}{(1-R)/p}.
\end{eqnarray*}
Thus
\begin{eqnarray*}
	NB = NBI\times \frac{PPV-R}{PPV-\pi}.
\end{eqnarray*}  
Thus $NBI<NB$ if $R<\pi$, $NBI>NB$ if $R>\pi$, and $NBI=NB=J\pi$ at $R=\pi$ ($J=Sens+Spec-1$ is Youden's index).

\section{Variance of MRS, and hypothesis testing for two MRSs}

\begin{table}
	\caption{Simulation of MRS, Youden's index, their standard errors (se), and 95\% confidence interval (CI) overage.}
	\begin{tabular}{|r|r|r|r|r|r|r|}
		\hline
		   \multicolumn{1}{|r|}{} & \multicolumn{2}{c|}{0.78\% threshold} & \multicolumn{2}{c|}{10\% threshold} & \multicolumn{2}{c|}{30\% threshold}  \\ \hline
		   \multicolumn{1}{|r|}{} & Youden & MRS   & Youden & MRS   & Youden & MRS \\
		\hline
		true parameter & 0.37587 & 0.016721 & 0.24422 & 0.010857 & 0.17878 & 0.007947 \\
		mean estimate & 0.3757 & 0.016711 & 0.24406 & 0.010858 & 0.17861 & 0.007934 \\
		empirical se & 0.03933 & 0.002354 & 0.04436 & 0.002217 & 0.03858 & 0.001862 \\
		estimated se & 0.03885 & 0.002343 & 0.04409 & 0.002209 & 0.03819 & 0.00185 \\
		95\% CI & 94.037 & 94.767 & 94.442 & 94.25 & 93.929 & 94.231 \\
		\hline
	\end{tabular}%
	
	\label{tab:simulation}%
\end{table}%

Asymptotic variances for MRS and Youden's index follow from applying the delta method to the quadrinomial variance matrix from a 2x2 table with $n$ as the sample size and $a=P(D+,M+),~b=P(D+,M-),~c=P(D-,M+),~d=P(D-,M-)$.  Each variance is $g^TVg$ where $g(a,b,c,d)$ is the usual gradient of the quantity.  $V$ is the usual quadrinomial variance matrix of the cell probabilities, for total sample size $n$: 

\[
V=1/n \times
\begin{bmatrix}
a(1-a) & -ab & -ac & -ad \\
-ba & b(1-b) & -bc & -bd \\
-ca & -cb & c(1-c) & -cd \\
-da & -db & -dc & d(1-d)
\end{bmatrix}
\]

Recall that MRS is twice the cross-product difference of joint probabilities inside a 2x2 table: $MRS = 2(ad-bc)$ (see main paper Appendix).  The variance of MRS is based on the gradient $g=2[d~,~-c~,~-b~,~a]$.  Calculating $g^TVg$ yields the variance
\begin{eqnarray*}
	Var(MRS) &=& 4\{ ad(a+d)+bc(b+c)-MRS^2 \}/n.
\end{eqnarray*}
This variance requires only the sample proportions of the joint probabilities.  It does not require fixed or \textit{a priori} known test positivity or disease prevalence. 

Table 1 examines the properties of the MRS variance and MRS confidence interval coverage by simulation.  We did 1 million simulations for each of 3 quadrinomial 2x2 tables based on the 3 cutpoints we considered in the Washington Ashkenazi Study (WAS) : 0.78\%, 10\%, and 30\%.  The quadrinomials were based on sample size of 4589 (as in WAS), with expectations for the $[a,b,c,d]$ cell counts as: 
\begin{enumerate}
	\item $[84.72, 19.73, 1951.88, 2532.67]$ for 0.78\%
	\item $[29.63, 74.75, 177.70, 4306.92]$ for 10\%
	\item $[19.74, 84.62, 46.52, 4438.11]$ for 30\%.
\end{enumerate}
In all cases, MRS and its variance were estimated with little bias, and 95\% confidence intervals performed nominally (Table 1).

To ensure proper MRS confidence intervals, note that $MRS\in [-0.5,0.5]$.  This is easy to see based on the MRS expression $MRS=2J\pi(1-\pi)$, where $J$ is Youden's index and $\pi$ is disease prevalence.  The maximum/minimum MRS of $\pm0.5$ occurs when $\pi=0.5$ and $J=\pm1$.  Thus $(0.5+MRS)\in [0,1]$ and a logit transformation of $(0.5+MRS)$ will ensure that confidence intervals are proper.  Applying the delta method yields
\begin{eqnarray*}
	Var(logit(0.5+MRS)) = ( (0.5+MRS)(0.5-MRS) )^{-2} \times Var(MRS).
\end{eqnarray*}
This variance is used to calculate confidence intervals on the $logit(0.5+MRS)$ scale.  Then, convert back to the MRS scale by applying to each endpoint $x$ the inverse function
\begin{eqnarray*}
	\frac{e^x}{1+e^x} - \frac{1}{2}.
\end{eqnarray*}

\subsection{Hypothesis testing if two MRSs are equal}

In general, testing if two independent MRSs differ can be based on the difference of the two MRSs, whose variance would be $Var(MRS_1)+Var(MRS_2)$.  But if the MRSs are calculated within the same population, and hence same prevalence, the ratio of MRSs is a better statistic.  This is because the nuisance parameter of prevalence cancels out, leaving the ratio of Youden's indices $J_1,J_2$:
\begin{eqnarray*}
	\frac{MRS_1}{MRS_2} = \frac{2J_1\pi(1-\pi)}{2J_2\pi(1-\pi)} = \frac{J_1}{J_2}.
\end{eqnarray*}
We will calculate the variance of the log of the ratio of two independent Youden's indices, based on a quadrinomial likelihood for 2x2 tables.  

Recall that Youden's index $J=Sens+Spec-1$ can be written in terms of the joint probabilities in a 2x2 table:
\begin{eqnarray*}
	J &=& \frac{a}{a+b} + \frac{d}{c+d} - 1.
\end{eqnarray*}
Then, to first calculate the variance of a single Youden's index, the gradient is
\[
g=
\begin{bmatrix}
\frac{b}{(a+b)^2} & \frac{-1}{a(1+b/a)^2} & \frac{-1}{d(1+c/d)^2} & \frac{c}{(c+d)^2} 
\end{bmatrix}
\]
The variance of a single Youden's index $J$ is $V_J=g^TVg$.  Table 1 shows that the variance for Youden's index is unbiased and 95\% confidence intervals performed nominally.

For two independent Youden's indices, the variance of the log of their ratio is asymptotically
\begin{eqnarray*}
	V_{12} &=& \frac{V_{J_1}}{J_1^2} + \frac{V_{J_2}}{J_2^2},
\end{eqnarray*}
and asymptotically $log(J_1/J_2)/\sqrt{V_{12}}\sim N(0,1)$.

The p-values comparing MRSs in sections~5.1 and~5.2 of the main paper are based on the ratio of Youden's indices.  For Ashkenazi-Jews, comparing the MRSs at a 0.78\% threshold vs. 10\%, the p-value based on the difference of MRSs is 0.0703, but that based on ratio of Youden's indices is 0.0392 (as reported in the main paper section 5.1).  For comparing the 0.78\% threshold vs. 30\%, the p-value based on the difference of MRSs is 0.0035, but that based on ratio of Youden's indices is 0.00054 (as reported in the main paper section 5.2).  In each situation, the smaller p-values by using the ratio of Youden's indices reflects the gain in statistical power by removing the nuisance parameter, disease prevalence $\pi$.

\section{Figure 1: Example cancer family history and family tree}

\begin{figure}[t!]
	\includegraphics[angle=-0,width=3in,]{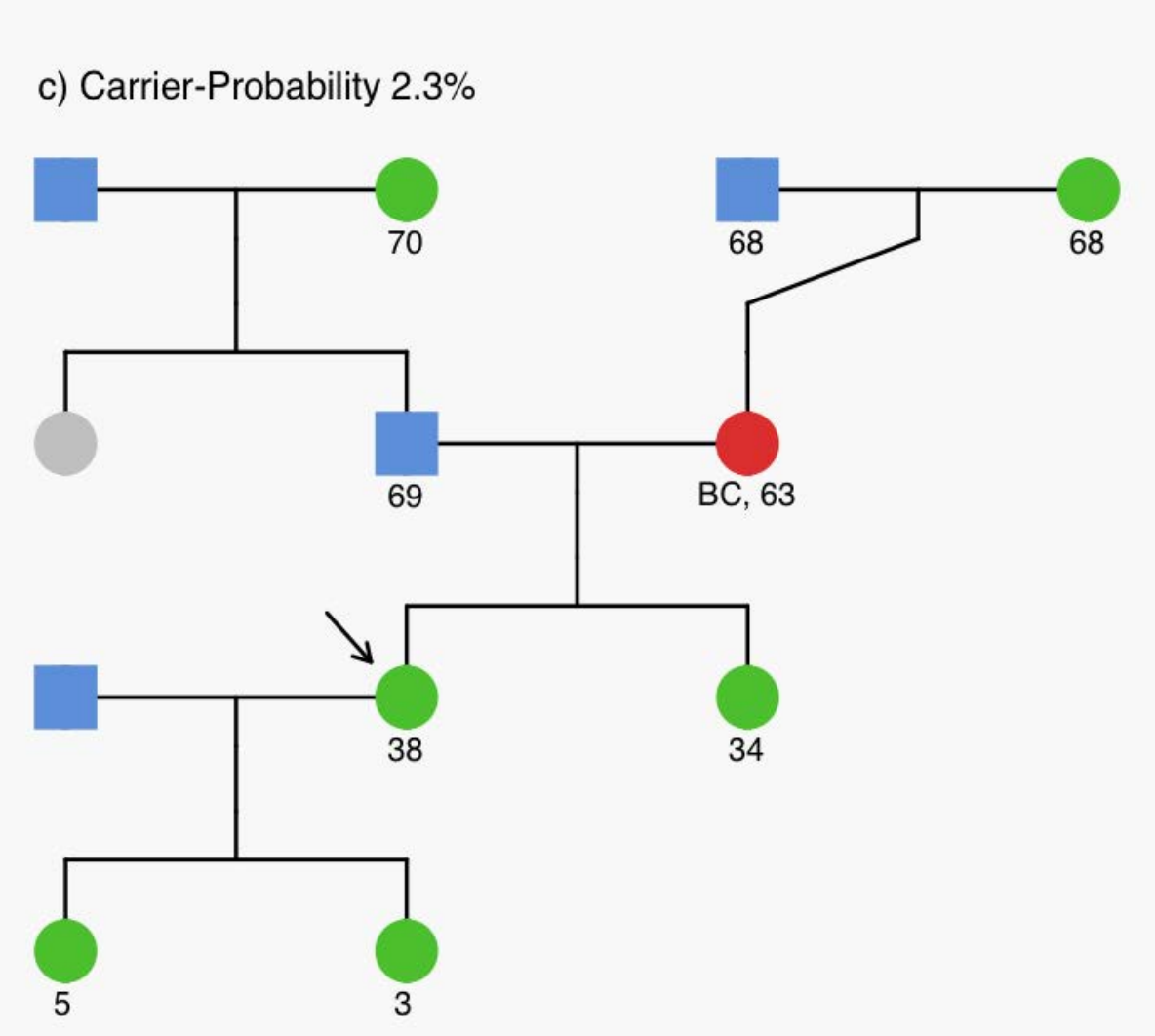}    
	\includegraphics[angle=-0,width=2in,]{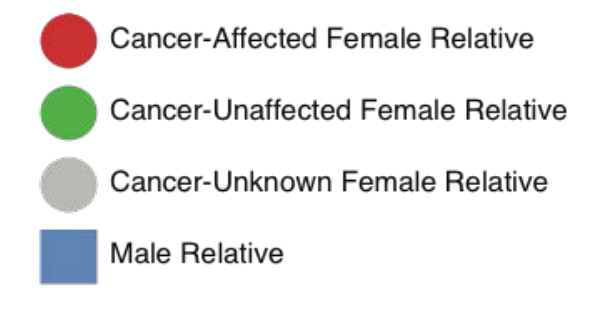}    
	\caption{Based on this pedigree, BRCAPRO calculates that the volunteer (arrow) has 2.3\% chance of carrying a \textit{BRCA1/2} mutation.  Mother had breast cancer at age 63, no other cancers in the family.}
	\label{fig:Pedigree}
\end{figure}

\section{Fixing the ROC curve to account for disease prevalence: The frequency-scaled ROC curve and its relationship of AUC and MRS}
\label{fROC}

The ROC plots $P(M+|D+)$ vs. $P(M+|D-)$.  The frequency-scaled ROC (fROC) plots $P(D+,M+)$ vs. $P(D+,M-)$~\citep{Hilden1991}.  Unlike the square ROC, fROC accounts for prevalence and is a $\pi$ by $1-\pi$ rectangle.  The diagonal is the uninformative fROC curve, the area under which is $\pi(1-\pi)/2$.  The area under the fROC curve for a test, which is a single point with lines extending to the bottom-left and top-right corners, can be shown to be $P(D+,M+)(1-\pi)/2 + P(D-,M-)\pi/2$.  The \textit{ratio} of the area under the fROC to the chance area for random selection, equals the AUC:
\begin{eqnarray*}
	\frac{area}{chance~area} &=& \frac{P(D+,M+)(1-\pi)/2 + P(D-,M-)\pi/2}{\pi(1-\pi)/2} = \frac{Sens+Spec}{2} = AUC.
\end{eqnarray*} 
The \textit{difference} between the the area under the fROC to the chance area equals $MRS/4$:
\begin{eqnarray*}
	area-chance~area = P(D+,M+)(1-\pi)/2 + P(D-,M-)\pi/2-\pi(1-\pi)/2 = MRS/4.
\end{eqnarray*} 
A high ratio (AUC) might conceal a small difference (MRS), which is apt to be the case for uncommon diseases.

\end{document}